\documentclass[
reprint,
showkeys,
longbibliography,
nofootinbib,
superscriptaddress,
aps,
notitlepage
]{revtex4-1}
\usepackage[utf8]{inputenc}

\usepackage{amsmath}
\usepackage{amssymb}
\usepackage{bm}
\usepackage{enumitem}
\usepackage[acronym]{glossaries}
\usepackage{graphicx}
\usepackage[colorlinks,unicode]{hyperref}
\usepackage[capitalise]{cleveref}  
\usepackage{mathtools}
\usepackage{multirow}
\usepackage{physics}
\usepackage{siunitx}
\usepackage[dvipsnames]{xcolor}
\usepackage{tcolorbox}

\DeclareSIUnit\week{week}

\makeatletter
\renewcommand\onecolumngrid{%
\do@columngrid{one}{\@ne}%
\def\set@footnotewidth{\onecolumngrid}%
\def\footnoterule{\kern-6pt\hrule width 1.5in\kern6pt}%
}
\makeatother

\newcommand\myshade{80}
\colorlet{mylinkcolor}{ForestGreen}
\colorlet{mycitecolor}{Red}
\colorlet{myurlcolor}{violet}

\hypersetup{
  linkcolor  = mylinkcolor!\myshade!black,
  citecolor  = mycitecolor!\myshade!black,
  urlcolor   = myurlcolor!\myshade!black,
  colorlinks = true
}

\newcommand{\calE}{\mathcal{E}}

\DeclareSymbolFont{mathtx}{OML}{txmi}{m}{it}
\DeclareMathAlphabet\mathbfcal{OMS}{cmsy}{b}{n}
\DeclareMathSymbol{v}{\mathalpha}{mathtx}{118}

\newcommand{\GRAPPA}{Gravitation Astroparticle Physics Amsterdam (GRAPPA),\\ University of Amsterdam, 1098 XH Amsterdam, The Netherlands}

\begin{document}

\title{Measuring the neutron star equation of state\\ from EMRIs in dark matter environments with LISA}

\author{Theophanes K. Karydas}
\email{t.karydas@uva.nl}
\affiliation{\GRAPPA}

\author{Gianfranco Bertone}
\affiliation{\GRAPPA}

\begin{abstract}
Gravitational-wave observations of extreme mass-ratio inspirals (EMRIs) in vacuum are largely insensitive to the internal structure of the small compact companion. We show that this conclusion can change when the central black hole is surrounded by a dense dark matter environment. We compute, for the first time, the relativistic dynamical-friction force on a neutron star moving through a collisionless medium and its impact on the evolution of EMRIs embedded in dense dark matter spikes. We then perform a Bayesian parameter-estimation analysis of simulated LISA observations to assess the measurability of both spike properties and the companion's internal structure.We find that, in our fiducial dark matter spike models, EMRIs with signal-to-noise ratio (SNR) $\gtrsim 120$ can already distinguish neutron star from black hole companions. At comparable signal-to-noise ratios, discrimination among neutron star equations of state begins to emerge and becomes significantly stronger toward the high-SNR tail.
\end{abstract}

\maketitle

\paragraph*{\bf Introduction.}
With the advent of next-generation millihertz gravitational-wave (GW) detectors \cite{LISA:2024hlh,TianQin:2015yph,Hu:2017mde}, it will become possible to monitor extreme mass ratio black hole (BH) binaries over long timescales. Such measurements will enable detailed investigations of the astrophysical environments where binaries form and evolve~\cite{Cardoso:2019rou, LISA:2022kgy, CanevaSantoro:2023aol, Bertone:2024rxe, Zwick:2025wkt, Zwick:2025wkt, Cardoso_2022, spieksma2025blackholespectroscopyenvironments, Cole:2022yzw, Roy:2024rhe}, opening a wide range of possibilities for fundamental physics. These range from stringent probes of gravity~\cite{Amaro_Seoane_2007, Gair_2013, speri2024probingfundamentalphysicsextreme, Mitra_2024} to searches for new physics in the particle sector~\cite{PhysRevLett.133.121404, Barack_2019, Bertone_2020, Baumann_2022, Maselli_2022, miller2025gravitationalwaveprobesparticle}. A particularly compelling scenario involves dark matter (DM) overdensities, or \emph{spikes}, around BHs~\cite{Gondolo_1999,Zhao_2005,Bertone:2005xz,Sadeghian_2013,Ferrer_2017,bertone2024darkmattermoundsrealistic,Ullio_2001}, which can influence the orbital evolution of inspiraling binaries~\cite{Eda_2013, Kavanagh_2020, Coogan_2022, Cole_2023, karydas2024sharpeningdarkmattersignature, kavanagh2024sharpeningdarkmattersignature, Speeney_2022, Becker_2022, mitra2025extrememassratioinspirals, Li_2022, Zhang_2024, Mukherjee_2024, Hannuksela_2020, Becker_2023, zhou2025intermediatemassratioinspiralsgeneraldynamical, Nichols_2023, Edwards_2020, Montalvo_2024, Vicente:2025gsg,karydas2025massspincoevolutionblack}.

The interaction of EMRIs with standard non-interacting particle DM is mediated only by gravity. The leading environmental effects in this case arise from dynamical friction \cite{chandra1, chandra2, chandra3} and, for BH companions, from particle accretion~ \cite{Misner:1973prb,Traykova_2023,karydas2024sharpeningdarkmattersignature,karydas2025massspincoevolutionblack}. Both effects are sensitive to the internal degrees of freedom of the small companion object, such as its mass and spin \cite{Dyson:2024qrq, Costa_2018, wang2024gravitationalmagnuseffectscalar,mach2025accretionvlasovgaskerr}.

Neutron stars (NS) are extremely compact astrophysical objects~\cite{Lattimer_2001,Lattimer_2004,chatziioannou2024neutronstarsdensematter}, and their interior provides a unique laboratory for the equation of state (EOS) of strongly interacting matter at supranuclear densities \cite{_zel_2016}, where exotic degrees of freedom \cite{PhysRevC.103.025808,Annala_2020,BRAMANTE20241,Zhang_2023} or phase transitions \cite{Dexheimer_2018} may occur. Probing this regime is one of the central goals of multi-messenger astronomy \cite{batista2021eucaptwhitepaperopportunities,Raaijmakers_2021,sathyaprakash2019multimessengeruniversegravitationalwaves}. Thus far, the detections of comparable mass ratio compact binaries by the LVK \cite{LIGO,VIRGO,KAGRA} collaboration have yielded constraints on NS properties \cite{PhysRevD.107.024021,PhysRevLett.119.161101,biswas2025simultaneouslyconstrainingneutronstar,Abbott_2020,Abbott_2018,Radice_2018,Raaijmakers_2020},  offering direct empirical insight into the behavior of dense matter. In sharp contrast, GWs from EMRIs in vacuum (with mass ratio, $m/M \leq 10^{-5}$) are often argued to be insensitive to the nature of the companion \cite{Pani_2019}.

Here, we demonstrate that environmental effects from DM can render EMRI waveforms sensitive to the nature of the small compact object. Specifically, we study the impact of a DM spike on the GW emission of an EMRI consisting of a NS orbiting a supermassive BH. 
We show that for our fiducial spike models (i), the spike parameters can be accurately recovered already for any detectable waveform, largely irrespective of the nature of the companion, and (ii) for sufficiently high SNR, the nature of the companion can be distinguished.\\

\paragraph*{\bf Setup.}
We consider an EMRI in which a NS orbits a massive BH embedded in a dark matter DM spike. As the companion moves through the surrounding DM distribution, dynamical friction provides an additional dissipative channel that alters the orbital evolution~\cite{Eda_2013,Kavanagh_2020}. To model the inspiral, we adopt the adiabatic framework\footnote{Adiabatic EMRI models have small systematic biases \cite{khalvati2025systematicerrorsfastrelativistic}, and including postadiabatic corrections may further improve parameter resolution \cite{burke2024accuracyrequirementsassessingimportance}, e.g. for secondary spin.}, in which the EMRI is described as a slow secular drift through a sequence of bound geodesics of the background BH spacetime~\cite{Khalvati_2025,Hughes_2005}.

At leading order, the secular evolution is governed by the orbit-averaged fluxes of energy, angular momentum, and Carter’s constant \cite{Misner:1973prb,PhysRev.174.1559}. In this work, we study quasi-circular inspirals around a non-spinning BH where due to symmetry inspirals are characterized only by the flux for angular momentum loss. Under these assumptions, we previously  derived the rate of change of the companion's angular momentum $L$ when at separation $r$ as\cite{Vicente:2025gsg},
\begin{equation} \label{eq:angtum}
    \frac{dL}{dt} = r F_3 \sqrt{1 - \frac{2GM}{c^2 r}}\,,
\end{equation}
where the root is a strong field correction (at most an $\sim\!18\%$ suppression at the innermost stable circular orbit with respect to the non-relativistic case, $\dot{L} = rF_3$) and $F_3$ is the dynamical friction force acting on the NS in the direction of its motion which we compute below. This flux will then be incorporated within the \texttt{FastEMRIWaveforms\_v2.0.0} (FEW) framework \cite{Katz:2021yft,Chua_2021,Speri_2024,chapmanbird2025fastframedraggingefficientwaveforms} to generate highly accurate GW waveforms.\\

\paragraph*{\bf Dynamical friction on neutron stars.}
\label{sec:dynamical_friction}
Dynamical friction is the dominant mechanism through which compact objects exchange momentum with their surrounding medium in collisionless environments such as cold DM spikes \cite{chandra1, chandra2, chandra3}. While Newtonian prescriptions exist for both point masses and extended perturbers \cite{extended,Morton_2025,gorkavenko2025dynamicalfrictionultralightdark}, and relativistic formulations have been developed for BHs \cite{Misner:1973prb,Traykova_2023}, NSs cannot be fully captured by either framework. This is because they possess a finite size, EOS-dependent density profiles, and an extreme compactness which makes it impossible to neglect strong-gravity effects in their vicinity. To address this, we develop a relativistic framework for computing dynamical friction on NSs. In absence of a dissipative coupling between baryonic and dark matter, the force experienced by NS is given by
\begin{equation} \label{eq:force}
	\bm{F} = \int \gamma \, \mu \,f(\mathbf{v}) \, v \,\sigma_{\rm scatter}(v) \, \mathbf{v} \, \dd^3 \mathbf{v}\,,
\end{equation}
where $\mathbf{v}$ is the relative velocity of the particles sampled from the distribution function $f(\mathbf{v})$ \cite{binney,Sadeghian_2013} in the NS's frame of reference, $\gamma$ is the Lorentz factor, and $\mu$ is the particles' rest mass. The term $\sigma_\mathrm{scatter}$ represents the cross-section for exchange of momentum via gravitational scattering. For spherically symmetric interactions (cf. discussions), it is
\begin{equation} \label{eq:sigmaNS}
    \sigma_\mathrm{scatter} = 4 \pi \int_0^{b_\mathrm{max}} b \cos^2 \chi \,\, \mathrm{d}b \,,
\end{equation}
where $b_\mathrm{max} = r \left(q/3\right)^{1/3}$ is the Hill radius, and $\chi$ is the azimuthal difference in the position of a particle between infinity and the point of closest approach. It is related to the deflection angle through $\theta_\mathrm{defl} = 2\chi -\pi$ (cf. \cite[Fig. 3.2]{binney}). Evaluating \cref{eq:sigmaNS} requires solving the geodesic equations governing particle motion in the NS spacetime whose computation is discussed in \cref{app:geodesic}. 

To assess the impact of the DM environment on the companion’s orbit, we numerically integrate \cref{eq:sigmaNS} for the NS spacetime and insert it into \cref{eq:force} and \cref{eq:angtum}, which are themselves integrated over the spike’s velocity distribution. In \cref{app:friction} we give present the technical details of the calculation. We assume the relativistic distribution function from Ref. \cite{Sadeghian_2013}, derived for the adiabatic growth of a non-spinning BH, and transform it to the NS frame, accounting for the central BH’s strong gravity as in \cite{Vicente:2025gsg}. To capture a representative range of possible NS structures, we consider three EOS highlighted in \cite{2022ApJ...926...75B} for being compatible with with tidal deformability constraints from GWs \cite{Abbott_2018} and heavy pulsar observations \cite{Antoniadis_2013}: MPA1 \cite{MUTHER1987469}, AP3 \cite{Akmal_1998} and SLy \cite{Haensel_2004}, which are matched with a crust as in \cite{Douchin_2001}. The EOS and resulting radial profiles of the change in friction are shown in \cref{fig:equations_of_state} 
alongside a collection of other representative models.

\begin{figure}[t!]
\centering
\hspace*{-0.08\columnwidth}
    \includegraphics[width=0.95\columnwidth]{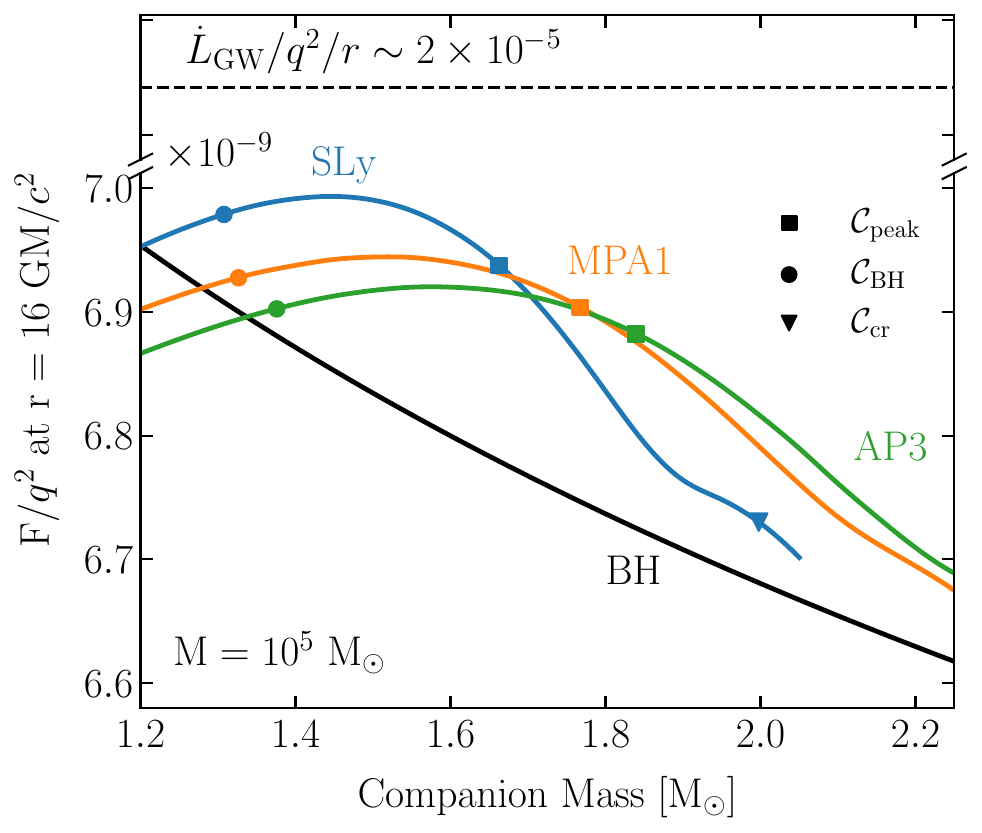}
    \caption{\textbf{Dynamical friction as a function of companion mass.} Markers indicate the predicted onset of key features from a constant-density–sphere toy model for $u = \langle u\rangle_{16GM/c^2}$. A NS’s crust causes the toy model to overestimate the true values, particularly for SLy's thick crust. The force is given for $c=G=1$.\label{fig:force_compactness}}
    \vspace*{-1em}
\end{figure}

In \cref{fig:force_compactness}, we show the drag force $F$ (normalised by $q^2$ so that the BH curve is approximately constant) as a function of the companion mass. The behaviour of the NS curves can be understood as follows. 
\begin{itemize}
    \item At low masses, the drag on NSs is weaker than for BHs due to their smaller compactness, which effectively dilutes the density probed by DM orbits and yields a smaller dynamical friction force than BH accretion. 
    \item As the mass (and hence the compactness) increases, the NS drag overtakes the BH value: this is a genuinely relativistic effect driven by nearly radial encounters that would have been accreted by a BH but are instead strongly deflected by the NS. Particles scattered through an obtuse angle transfer more momentum to the star than accretion would (cf.~\cite[Ch.~3]{binney}); in the extreme case of a $\sim 180^\circ$ deflection, the momentum transfer is equivalent to accreting two particles, effectively enhancing the cross-section. 
    \item For even larger masses, the force reaches a maximum and then decreases, because the higher compactness allows particles to wind around the star and exit with acute scattering angles, which are less efficient at transferring momentum and thus reduce the scattering cross-section. 
\end{itemize}

In \cref{app:toymodel_threshold}, we introduce an analytical toy model that reproduces this qualitative behaviour, and we use a constant-density sphere to estimate the compactness values at which the onset and peak of the enhancement occur.

Finally, while low-compactness stars may allow particles to graze their surface and smoothly reach deeper with decreasing impact parameter, sufficiently compact objects prohibit such grazing trajectories. In this case, radial encounters with $b < b_\mathrm{cr}$ penetrate the interior whilst effectively on a plunging trajectory, and reach discontinuously deep within the star before escaping due to the decrease in enclosed mass at the core. These will wind discontinuously more compared to slightly less radial encounters, thus changing the scattering cross-section for those compact objects. This effect arises as a generalization to the photon-sphere to account beyond just photons but particles moving with any velocity. Amongst our selected models, this only affects SLy which achieves the appropriate compactness.\\

\paragraph*{\bf Parameter estimation with incorrect models.}

Having established the physical setup and the modeling framework for computing NS inspirals within DM spikes, we now turn to assessing the observability of these effects through future GW observations in the mHz band such as with LISA \cite{LISA:2022kgy}. In particular, we first explore how parameter estimation is affected when the analysis is carried out with an incorrect model for the companion, and quantify the resulting biases in the inferred system parameters.

We quantify these biases using the Bayesian framework described in \cref{app:bayesian}. Specifically, we inject a waveform with a duration of 1 year, emitted by a binary with a NS companion described by the SLy equation of state and the intrinsic parameters of a benchmark system \linebreak ($M=10^5~\mathrm{M}_\odot$, $m = 2~\mathrm{M}_\odot$, and $\rho = 10\,\rho_\mathrm{DS}\approx 3\times10^{18}~\mathrm{M}_\odot/\mathrm{pc}^3$ at $10^{-6}~\mathrm{pc}$, where $\rho_\mathrm{DS}$ represents the expected density for a Draco dSph-like galaxy~\cite{Wanders_2015}) and then recover it using the true model or three alternative companion models: MPA1, AP3, and a BH. 

EMRIs are expected to achieve SNRs up to $10^3$ \cite{Speri_2024,Babak_2017}, which when rescaled to the lower primary and companion masses considered here under the assumption that largest SNR's originate from the LISA sensitivity peak \cite{Khalvati_2025}, corresponds to an SNR $\sim 400$ ($D_L \sim 22~\mathrm{Mpc}$). We adopt an SNR of $200$ ($D_L \sim 44~\mathrm{Mpc}$) as a fiducial benchmark, chosen to probe the high-SNR tail of the event distribution, where model selection is most informative, rather than to represent a typical source. We emphasize, however, that EMRI rate predictions span nearly three orders of magnitude~\cite{Babak_2017,Mancieri_2025,Mazzolari_2022}, and that estimates for systems with NS secondaries have so far been carried out only for in-situ formation channels~\cite{Aharon_2016}. The density and SNR choices adopted here should therefore be regarded as convenient benchmarks, rather than forecasts.

\begin{figure}[t!]
\centering
\hspace*{-0.08\columnwidth}
    \includegraphics[width=\columnwidth]{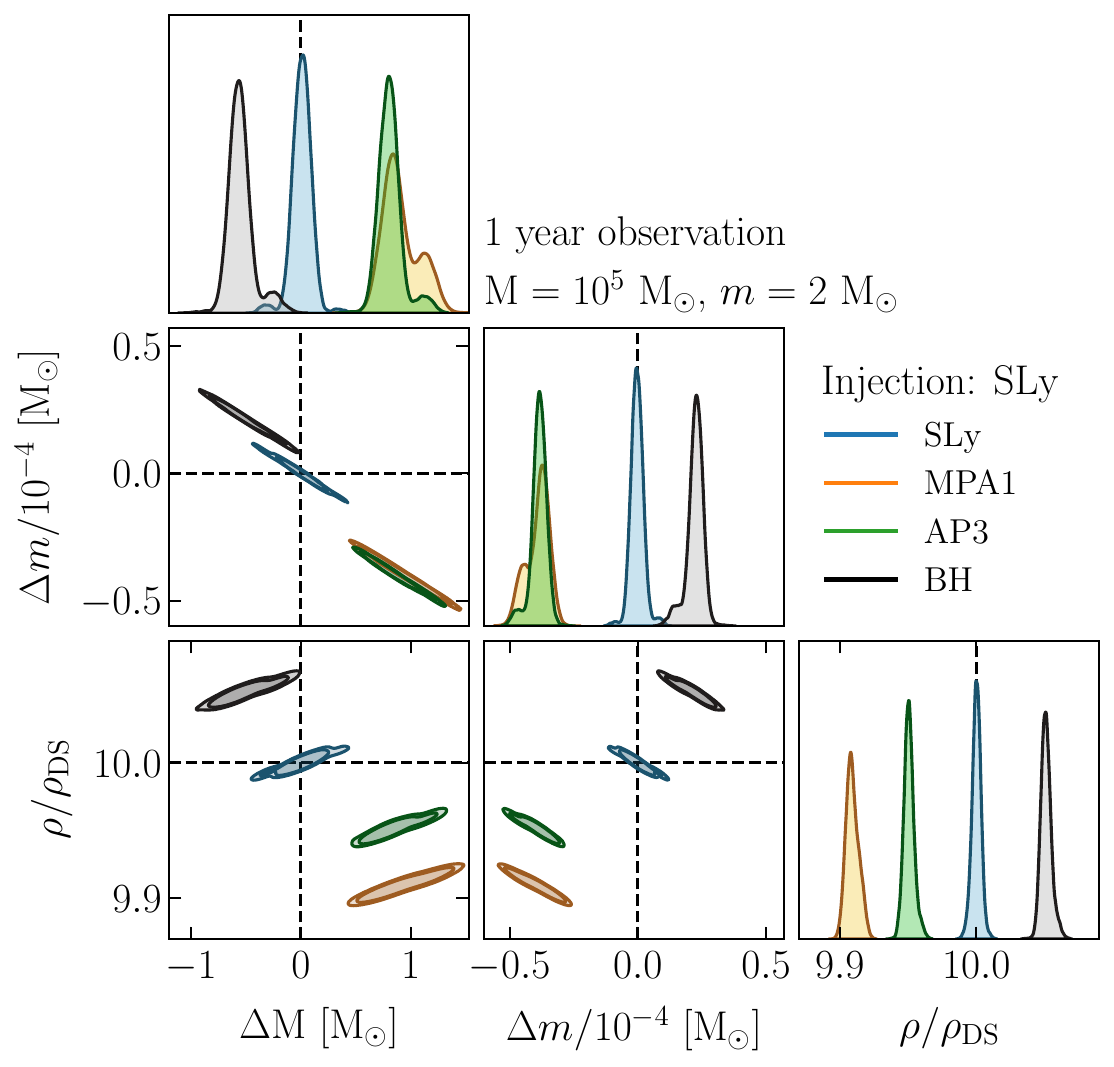}
    \caption{\textbf{Marginal posteriors for the masses and spike density.} Results are obtained with 1 year of observation data at SNR = 200, with black dashed lines indicating the injected values. \label{fig:corern_ns}}
    \vspace*{-1em}
\end{figure}

In \cref{fig:corern_ns} we show the resulting one- and two-dimensional marginal posteriors, illustrating how the choice of companion model affects parameter recovery. We find that the component masses and DM spike normalization are recovered with similar accuracy across all companion models, with only a modest bias when an incorrect model is used. In particular, the $\sim 1\%$ bias in the inferred DM density, shows that spikes remain clearly identifiable even when the companion assumption is incorrect, which highlights the potential of LISA to probe DM spikes.\\

\paragraph*{\bf Inferring the equation of state.}
We now turn to the prospects for inferring the NS EOS from GW observations of EMRIs in DM spikes. For each of the three NS models (SLy, MPA1, AP3), we generate mock data sets containing an EMRI signal that includes the contribution of dynamical friction, and analyse each injection under four companion hypotheses: SLy, MPA1, AP3, and a BH. For every case, we compute the Bayesian evidence and construct Bayes factors, providing a quantitative measure of how strongly the data support each model and repeat for a range of SNRs $\in \left[200, 400\right]$. This allows us to assess whether the different companion models can be distinguished across the NS mass spectrum.

\begin{figure*}[t!]
    \centering
    \vspace*{-1em}
    \hspace*{-0.08\columnwidth}
    \includegraphics[width=\textwidth]{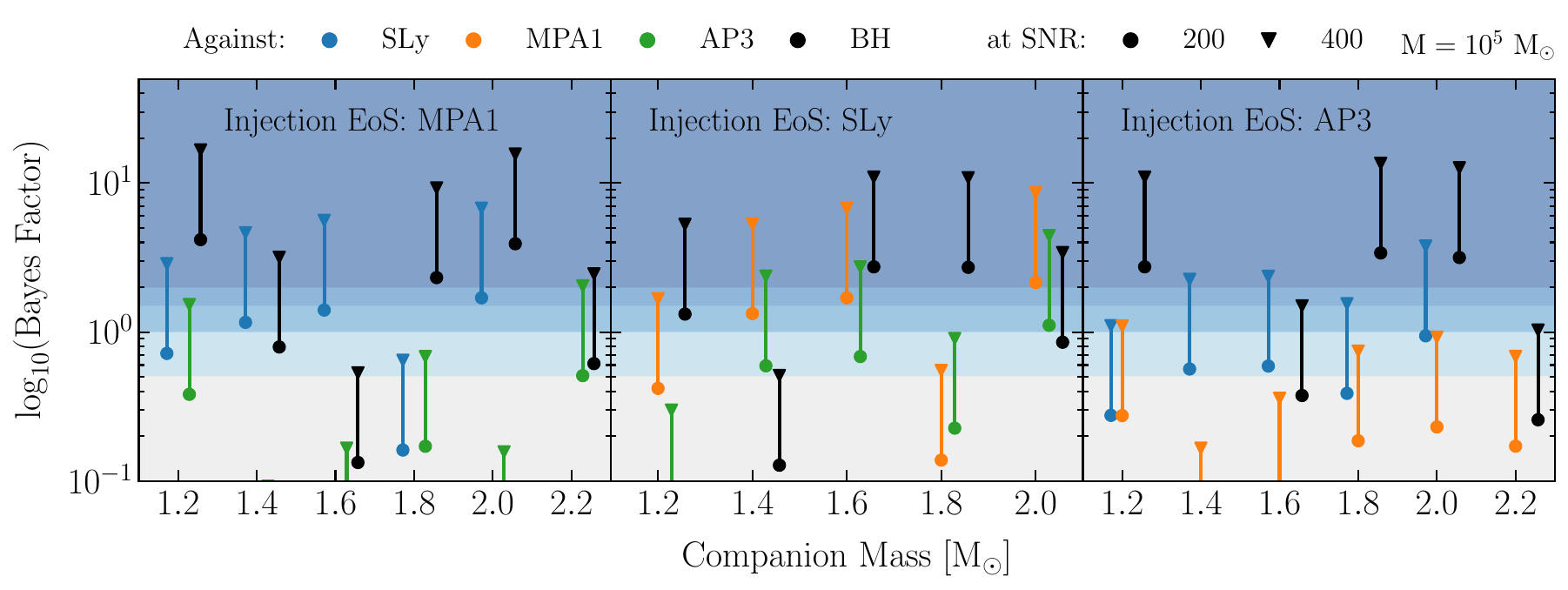}
    \vspace*{-1em}
    \caption{\textbf{Bayes factors for EOS comparisons with different companion masses.} Colored bands denote the Jeffreys-scale interpretation \cite{Jeffreys1939-JEFTOP-5} from ``Trivial'' (gray), to ``Substantial'', ``Strong'', ``Very Strong'', and ``Decisive'' (dark blue). Errors obtained through the nested-sampling evidence uncertainty would only be visible for the systems in the trivial regime and are therefore omitted. The SLy EOS cannot achieve a $2.2~\mathrm{M}_\odot$ configuration and is hence omitted.\label{fig:bayes_factors}}
\end{figure*}

The results of this analysis are summarised in \cref{fig:bayes_factors}, which shows the Bayes factors $\mathcal{B}$ for comparisons between different companion models across the range of NS masses considered. Most comparisons yield strong evidence ($\mathcal{B} > 10$) in favour of the true model, although some pairs provide less than substantial support ($\mathcal{B} < 3.2$), primarily for MPA1 versus AP3, which produce very similar NSs. In a subset of cases the Bayes factors reach the regime of decisive evidence ($\mathcal{B} > 100$). This is remarkable given that the EOS models we consider are already relatively similar, and all remain compatible with current tidal-deformability constraints from GW observations of binary NS mergers~\cite{Abbott_2018}. This highlights how DM environmental effects may allow us to probe the internal structure of the compact companion.

The mass dependence of the Bayes factors exhibits a correlation with the mass trends of the force.
In particular, we observe that the Bayes factors disfavoring a model, for example, a BH, tend to reach their largest (or smallest) values at specific masses, consistent with the distance between spatial gradients of the forces from those models.
Nevertheless, the shift is small, and the behavior at $r = 16~\mathrm{GM}/\mathrm{c}^2$ provides a particularly useful diagnostic, as the majority of the inspiral’s impact is accumulated at larger separations and that radius approximately corresponds to the furthest distance over the 1-year span of waveform data.

As expected, since $\ln \mathcal{B} \propto \mathrm{SNR}^2$ \cite{PhysRevD.103.062002,Cornish_2011}, it is the tail of high-SNR events that enables the most stringent tests, reaching Bayes factors up to $\mathcal{B} \sim 10^{20}$ in favor of the true model for $\mathrm{SNR}=400$.\\

\paragraph*{\bf Detectability Across Parameter Space.}
So far, we have demonstrated that the internal structure of the companion can be distinguished in a high-SNR EMRI embedded in a DM environment. We now turn to the conditions under which this effect remains detectable as we vary the system parameters. To this end, we use the Laplace approximation to the Bayesian evidence (cf.~\cref{app:bayesian}) to make the exploration of parameter space computationally tractable. 

\begin{figure}[t]
    \centering
    \hspace*{0.01\columnwidth}
    \includegraphics[width=0.95\columnwidth]{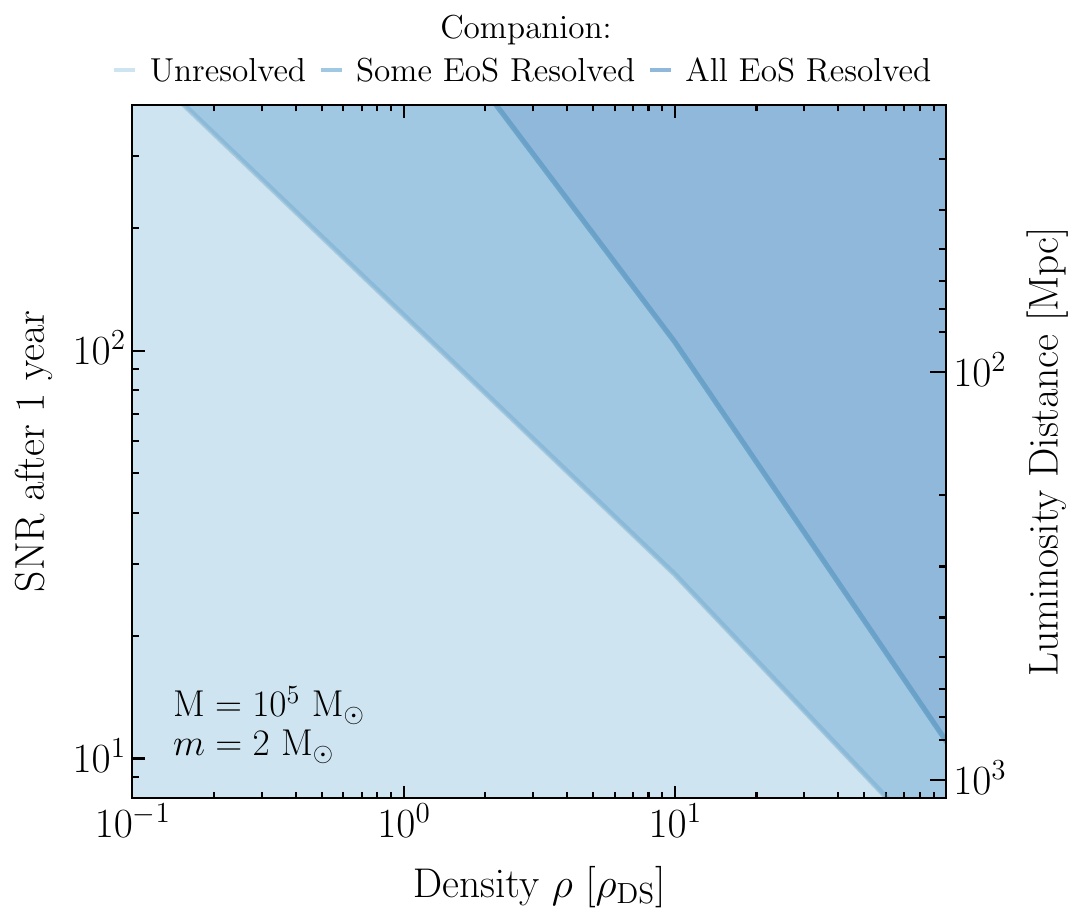}
    \vspace*{-1.2em}
    \caption{\textbf{Minimum optimal signal-to-noise ratio accumulated after 1 year for distinguishing EMRIs with neutron star companions.}
    Arrows point the SNR for binaries placed at $1~\mathrm{Gpc}$ distance.
    \label{fig:bands}}
    \vspace*{-1.5em}
\end{figure}

In \cref{fig:bands}, we illustrate how the discriminating power depends jointly on the SNR, the DM density, and the primary mass. The lines show the combinations of DM density and SNR for which the Bayes factor between the relevant models reaches the threshold for decisive evidence, $\mathcal{B}=100$. Reading the plot
from low to high density or SNR, three regimes emerge.
At low density/low SNR, the presence of a spike can be established with decisive evidence until $\mathrm{SNR}\approx 120$ for a $10^5\,\mathrm{M}_\odot$ primary, while the companion models remain degenerate. At higher SNR, subsets of EOS models yield $\mathcal{B}\ge 100$, and in the highest-density/highest-SNR regime ($\mathrm{SNR}\gtrsim 800$) all of our EOS comparisons satisfy the same decisive-evidence criterion.

\paragraph*{\bf Discussion.}
Ou analysis can be applied to other astrophysical objects, like main sequence stars ($\mathcal{C} \sim 10^{-6}$) or white dwarfs ($\mathcal{C} \sim 10^{-3}$) within environments with relativistic velocity distributions. It can also be applied to exotic compact objects like strange quark stars, composed of deconfined strange quark matter \cite{1986ApJ...310..261A,1986A&A...160..121H} which are generally more compact than ordinary NS; or boson stars \cite{JETZER1992163}, which could form from ultralight or self-interacting scalar fields \cite{PhysRevD.97.063012}. Additional targets can be horizonless BH mimickers (cf. \cite{bambi2025blackholemimickerstheory} and references therein), gravastars \cite{gravastar}, or Q-balls \cite{Coleman:1985ki}, and even BHs with small environments around them, such as gravitational atoms \cite{Arvanitaki_2015} formed when ultra-light bosonic fields grow through superradiance \cite{Brito_2020}, though one must adapt the formalism beyond spherical symmetry for most atom configurations.

Although the most striking implication of our analysis is the
possibility of probing the internal structure of the companion,
the region of parameter space in which LISA can distinguish
NS from BH companions is itself of
astrophysical interest. The detection of unusually low-mass BHs, in particular, would point to nonstandard formation channels: they could, for instance, be primordial BHs \cite{Carr:2020gox,Green_2021}, or BHs formed through the DM–induced collapse of NSs. DM–nucleon interactions can, in fact, lead to accumulation of DM inside a NS. If the captured DM becomes self-gravitating and collapses to form a BH, the latter can grow by accreting nuclear matter and eventually consume the star from the inside (see, e.g., Refs.~\cite{Goldman:1989nd,Bertone:2007ae}). 

 Accounting for DM-nucleon interactions, if present, in the complementary regime in which the DM mass and cross section do not trigger collapse over the NS lifetime, would require adapting the deflection formalism to include nuclear scattering. The extreme case where all particles crossing the interior are captured, the dynamical friction induced by a NS would be analogous to that induced by a BH, but with a larger geometric cross-section. We find that for a low NS compactness ($\mathcal{C} \sim 0.1$) it can be at most $\sim 50\%$.
  Therefore, realistic couplings\footnote{Requiring the DM mean free path $\sim m_n/(\rho_\mathrm{nucleus} \sigma_{\chi n})$, to exceed the NS radius so that particles undergo multiple scatterings before becoming gravitationally bound implies $\sigma_{\chi n} \gg 10^{-44}~\mathrm{cm}^2$. Current limits already constrain $\sigma_{\chi n} \lesssim 10^{-41}~\mathrm{cm}^2$ \cite{Aprile_2023}, greatly restricting the viable parameter space for this extreme scenario.} would only provide a subdominant contribution to the dynamical-friction force, highlighting the robustness of our analysis against possible DM–nucleon couplings.

We have neglected the spin of the NS, as its inclusion would affect our results only at higher order. Spin modifies the stellar structure \cite{1968ApJ...153..807H,c3jx-5487} and, at sufficiently rapid rotation, leads to mass shedding \cite{1992PhRvD..46.4161G}; it also alters the exterior space-time from Schwarzschild but to a less pronounced extent compared to Kerr for the same spin \cite{Berti_2004,1968ApJ...153..807H,1994ApJ...424..823C}. Kerr BHs can therefore be used to obtain an upper bound on spin-induced modifications to dynamical friction. The effect scales as $\tilde{a}^2$ of the dimensionless spin, $\tilde{a} = cJ/Gm^2$ rather than only with angular momentum $J$, and changes at most by $\sim10\%$ for an extremal BH, $\tilde{a} \sim 1$, (cf. \cite{Dyson:2024qrq,wang2024gravitationalmagnuseffectscalar,karydas2025massspincoevolutionblack}), whereas physically motivated EOSs yield NS spins no larger than $\tilde{a} \simeq 0.7$ \cite{Berti_2004}. Furthermore, isolated NSs at birth typically have $\tilde{a} < 0.04$, and accretion can increase their spin only up to $\tilde{a} \lesssim 0.4$ \cite{Baub_ck_2013,1982Natur.300..728A}, consistent with the fastest-rotating observed pulsar J1748-2446ad \cite{2006Sci...311.1901H}. In fact, fewer than $1\%$ of NSs are observed to exceed $\tilde{a} > 0.05$ \cite{Du_2024}. To incorporate spin, the scattering framework would need to be extended following \cite{Gonzo_2023}, but applied to the spacetime sourced by a rotating NS, and its structure needs to be modified accordingly.

Our analysis focuses on primaries of $10^5\,M_\odot$.
However, the EMRI rate distribution extends to primaries up to $\sim 10^6\,M_\odot$
\cite{Mazzolari_2022,Babak_2017}. Repeating the analysis for
these larger masses, with the dark matter density normalized to
that of a Milky Way-like galaxy \cite{Gondolo_1999}, we find that
the ability to discriminate among companion models deteriorates
significantly. That's because smaller mass
ratios imply, for a fixed merger time, smaller initial
separations in units of the primary gravitational radius.
Consequently, the inspiral probes a narrower portion of the
dynamical friction profile, making features such as changes in slope more difficult to resolve.

We treated the spike as static, an approximation that is well controlled in the regime of interest. The feedback of the inspiralling companion on the spike was first studied in \cite{Kavanagh_2020}, and developed further in \cite{Coogan_2022,kavanagh2024sharpeningdarkmattersignature,karydas2024sharpeningdarkmattersignature}. 
Using the same semi-analytic framework, we estimate that for the mass ratio and orbital separations relevant here, the associated reduction in the dynamical friction force is below the $10^{-3}$ level, far smaller than the effect studied here.

The dark matter profile depends in general on the prior evolution of the host system.
Scattering off a stellar background can in principle soften dark matter cusps and spikes over long timescales \cite{Bertone_2005,Shapiro_2022}, but 
in the region probed by the gravitational-wave signal, the stellar density is so low that no star is expected within the relevant volume.  In realistic BH formation scenarios involving the collapse of supermassive stars, shallower dark matter ``mounds'' may form 
\cite{bertone2024darkmattermoundsrealistic}. 
Similarly, mergers can reduce the dark matter density \cite{Bertone:2024rxe}. In both cases, however, subsequent gas inflow, accretion, and regrowth is expected to rebuild the central overdensity \cite{Caiozzo_2026}.

For these reasons, Fig.~4 explores a range of dark matter density normalizations rather than relying on a single fiducial profile. A fully dynamical treatment of the dark matter distribution could shift the region of parameter space in which the effect is observable, but it does not alter the main conclusion: sufficiently dense dark matter environments can encode information about the companion's internal structure in the EMRI waveform.\\

\paragraph*{\bf Conclusions.} \label{sec:conclusions}

In this work, we have shown that 
DM overdensities around BHs, such as DM spikes or mounds around massive BHs \cite{Gondolo_1999, Sadeghian_2013, Ferrer_2017, bertone2024darkmattermoundsrealistic}, can render EMRI waveforms sensitive to the nature of the small compact object. In particular, we have explored the potential to probe NS internal structure and their equation of state \cite{Lattimer_2001,Lattimer_2004,chatziioannou2024neutronstarsdensematter} through EMRIs embedded within DM spikes observed by LISA \cite{LISA:2024hlh}.

We developed a relativistic framework for dynamical friction on compact objects moving through a collisionless medium, applicable to generic NS models with an arbitrary EOS as well as exotic compact objects, provided the internal metric is specified. Applying this framework to EMRIs with a NS secondary embedded in DM spikes, we computed the influence of the companion’s internal structure on the emitted gravitational waveforms using the adiabatic \texttt{FEW} framework \cite{Katz:2021yft,Chua_2021,Speri_2024,chapmanbird2025fastframedraggingefficientwaveforms}.

We find that, for our fiducial spike models, the binary masses and DM spike parameters can be accurately recovered for any detectable waveform, largely independently of the companion model. This implies that parameter estimation pipelines can be simplified without significant loss of precision. For sufficiently high signal-to-noise ratios, the additional friction induced by the companion’s internal structure becomes measurable. Using Bayesian model comparison, we have shown that such EMRIs can discriminate between NS and BH companions, and between some equations of state already within a year, if $\mathrm{SNR} \gtrsim 120$.
These results highlight that LISA observations of EMRIs embedded in a DM environment could serve as a new astrophysical laboratory, probing both the distribution of dark matter around massive BHs and the microphysics of matter at supranuclear densities.

\section*{Acknowledgments}
We thank Profs. Anna Watts and Charalampos Moustakidis for valuable discussions regarding neutron star physics, and Drs. Rodrigo Vicente and Violetta Sagun for feedback. We gratefully acknowledge the support of the Dutch Research Council (NWO) through an Open Competition Domain Science-M grant, project number OCENW.M.21.375.

\appendix

\section{Geodesic motion around neutron stars} \label{app:geodesic}

We consider a static, spherically symmetric and spinless NS in hydrostatic equilibrium. Under these assumptions, the stellar structure is obtained by solving the Tolman–Oppenheimer–Volkoff equations \cite{Misner:1973prb}, 
\begin{align}
    \frac{\mathrm{d}P}{\mathrm{d}r} &= - \frac{G m_r}{r^2 c^2} \left( \epsilon +P \right) \left( 1 +\frac{4\pi r^3 P}{m_r c^2} \right) \left( 1 -\frac{2 G m_r}{c^2 r} \right)^{-1}\!\!,\nonumber \\ 
    \frac{\mathrm{d}m_r}{\mathrm{d}r} &= 4 \pi r^2 \frac{\epsilon}{c^2}\,,
\end{align}
where $P(r)$, $m_r(r)$, $\epsilon(r) = \rho(r) c^2$ are the pressure, enclosed mass and energy density at radius $r$ respectively. The system is then solved given an EOS \cite{_zel_2016} $P(\epsilon)$ and by setting the core pressure/density. The latter is iteratively modified until a stable NS \cite{2000csnp.conf.....G} with the target total mass is produced.

Within the neutron star, the space-time is described by the metric \cite{Misner:1973prb},
\begin{equation} \label{eq:ns_metric}
    \mathrm{d}s^2 = - \tilde{g}_\mathrm{tt}(r) \, c^2 \mathrm{d}t^2 
    + \left(1 -\frac{2Gm_r}{r c^2}\right)^{-1} \!\! \mathrm{d}r^2 
    + r^2 \mathrm{d} \Omega^2 \,,
\end{equation}
where the term $\tilde{g}_\mathrm{tt}(r)$ is determined from the expression
\begin{equation} \label{eq:dlngtt}
    \frac{\mathrm{d} \ln \tilde{g}_\mathrm{tt}}{\mathrm{d}r} 
    = - \frac{2}{\epsilon + P} \frac{\mathrm{d}P}{\mathrm{d}r}\,,
\end{equation}
and the boundary condition $\tilde{g}_\mathrm{tt}(R) = 1 - 2Gm/(c^2 R)$. Outside the star, the space-time is taken to be Schwarzschild.

The orbits of massive particles are described by the Lagrangian $\mathcal{L} = -g_{\mu\nu} \dot{x}^\mu \dot{x}^\nu$, where $g_{\mu\nu}$ are the metric contributions to the differentials $dx^\mu dx^\nu$. Standard Lagrangian analysis yields the corresponding geodesic equations of motion,
\begin{align}
    \left(\frac{\mathrm{d}r}{\mathrm{d}\tau}\right)^2 \!\! &= \frac{1}{r^2} \!\left( 1 - \frac{2G m_r}{r c^2}\right) \!\left( \frac{\calE^2 r^2}{\tilde{g}_{tt} c^2} -h_z^2 -c^2 r^2 \right)\!, \label{eq:ns_drdt} \\
    \left(\frac{\mathrm{d}r}{\mathrm{d}\phi}\right)^2 \!\! &= r^2\! \left( 1 - \frac{2G m_r}{r c^2}\right) \! \left( \frac{\calE^2 r^2}{h^2_z \tilde{g}_{tt} c^2} -\frac{c^2 r^2 }{h_z^2} -1 \right)\!, \label{eq:ns_drdphi}
\end{align}
where the constants $\calE$, $h_z$ represent the specific energy and angular momentum, which are obtained by exploiting time-translation and axial symmetry of \cref{eq:ns_metric}.

For an incoming particle in the NS frame with impact parameter $b$ and velocity $u$ at infinity, the constants of motion  are: $\calE = \gamma c^2$, $h_z = \gamma b u$. Then, the angle $\chi$, drawn by an incoming particle until it reaches its closest approach $r_0$, is computed by integrating \cref{eq:ns_drdphi},
\begin{equation} \label{eq:chi_integral}
    \chi = h_z \int_{r_0}^{\infty} \frac{\left(1 - 2Gm_r/\left(rc^2\right)\right)^{-1/2}}{r \sqrt{\calE^2 r^2/\left(\tilde{g}_{tt}c^2\right)-c^2r^2 -h_z^2}} \, \mathrm{d}r \,,
\end{equation}
where $r_0$ is obtained at the root of $dr/d\tau$.

\section{Calculating dynamical friction} \label{app:friction}
If the point of closest approach is larger than the stellar radius $R$, particles follow geodesics entirely defined from the equivalent Schwarzschild spacetime as for a BH with the same mass. As such, we calculate the scattering cross-section only for those orbits that can probe the NS interior. The difference is given by,
\begin{equation} \label{eq:dsigma}
    \delta \sigma \equiv 4\pi \left[\int_0^{b_\mathrm{th}} \!b \cos^2 \chi_\mathrm{NS} \, \mathrm{d}b - \!\int_{b_\mathrm{cr}}^{b_\mathrm{th}} \!b \cos^2 \chi_\mathrm{BH} \, \mathrm{d}b\right],
\end{equation}
where $b_\mathrm{cr}$ is the critical impact parameter that separates accretion from deflection on BH geometry \cite{Misner:1973prb}, $b_\mathrm{th}$ is a threshold impact parameter below for which particles can penetrate the NS, and $\chi_\mathrm{NS/BH}$ are, respectively, the angles drawn by a particle in NS or BH spacetime. Using $\delta\sigma$ on \cref{eq:force}, we define
\begin{equation} \label{eq:forceNS}
    \bm{F}_\mathrm{NS} = \bm{F}_\mathrm{BH} +\delta \bm{F}\,,
\end{equation}
where $F_\mathrm{BH}$ is the dynamical friction experienced by a BH. In this work, we use the fit given by \citet{Traykova_2023} for relativistic dynamical friction on BHs.

We calculate $\delta \sigma$ by substituting \cref{eq:chi_integral} for the NS and BH spacetimes within \cref{eq:dsigma} and integrating in impact parameter. For the BH spacetime, we substitute $m_r(r) = m$ and $\tilde{g}_{tt} = 1 -2Gm/\left(rc^2\right)$. For less compact stars, the impact parameter $b_\mathrm{th}$ in the integration limit is determined by requiring the turning point of the orbit to coincide with the stellar radius. This condition is obtained by setting $dr/d\tau = 0$ at $r = R$ in the Schwarzschild metric, yielding
\begin{equation} \label{eq:bR}
    b_R^2 = R^2 \left( 1 +\frac{2 G m}{R c^2 \gamma^2 \beta^2}\right) \left( 1 - \frac{2 G m}{R c^2}\right)^{-1}\!\!\!.
\end{equation}
This reduces to the well-known Newtonian relation \cite{Landau1976Mechanics} in the non-relativistic limit $c \to \infty$. Note, however, that substituting $b_\mathrm{cr}$ into \cref{eq:ns_drdt}, it becomes apparent that around a BH, the distance of closest approach for massive particles cannot be smaller than
\begin{equation} \label{eq:r0min}
    r_0^\mathrm{min}(\beta) = \frac{Gm}{c^2} \frac{3\gamma^2 -4 +3\gamma\sqrt{\gamma^2 -8/9}}{2\left( \gamma^2 -1\right)}\,.
\end{equation}
This also suggests that particles of velocity $\beta$ cannot graze the surface of a very compact NS and escape. In particular, since $\lim_{\beta\to0}r_0^\mathrm{min} = 4Gm/c^2$, only NSs with compactness $\mathcal{C} \geq 0.25$ can be affected.\footnote{The effect can appear at smaller compactness if the star has a thick crust, which contributes disproportionately more to its radius than to its enclosed mass. The required condition is that \cref{eq:ns_drdt} exhibits a local minimum and maximum within $r_0^\mathrm{min}$, or equivalently that
\begin{equation}
    \frac{\mathrm{d}P}{\mathrm{d}r} = \frac{\rho c^2 +P}{r} \left( 1 -\frac{c^4}{\calE^2} \tilde{g}_\mathrm{tt} \right)\,,
\end{equation}
has two distinct and positive roots.} For these stars, radial particles with $b< b_\mathrm{cr}$ can still penetrate the interior, but the distance of closest approach undergoes a discontinuous jump, from $r_0^\mathrm{min}(\beta)$ to a point deep within the star when \cref{eq:ns_drdt} bifurcates. Physically, these particles are effectively on plunging trajectories and approach deep within the core but eventually escape due to the decrease in enclosed mass. In practice, this means that 
\begin{equation}
    b_\mathrm{th}(\beta,m,R) = 
    \begin{cases}
        b_R(\beta, m, R) \qquad &R \geq r_0^\mathrm{min}(\beta,m)\,, \\
        b_\mathrm{cr}(\beta, m) \qquad &\text{otherwise}\,, \\
    \end{cases}
\end{equation}
as long as $b_\mathrm{th} < b_\mathrm{max}$.

\section{The compactness dependence} \label{app:toymodel_threshold}

In this section, we provide an analytical argument to understand the observed relation between dynamical friction and NS mass seen in \cref{fig:force_compactness}. To factor out the effect of the velocity distribution we examine the cross-section at fixed velocity.

\begin{figure}[t!]
\centering
\hspace*{-0.08\columnwidth}
    \includegraphics[width=\columnwidth]{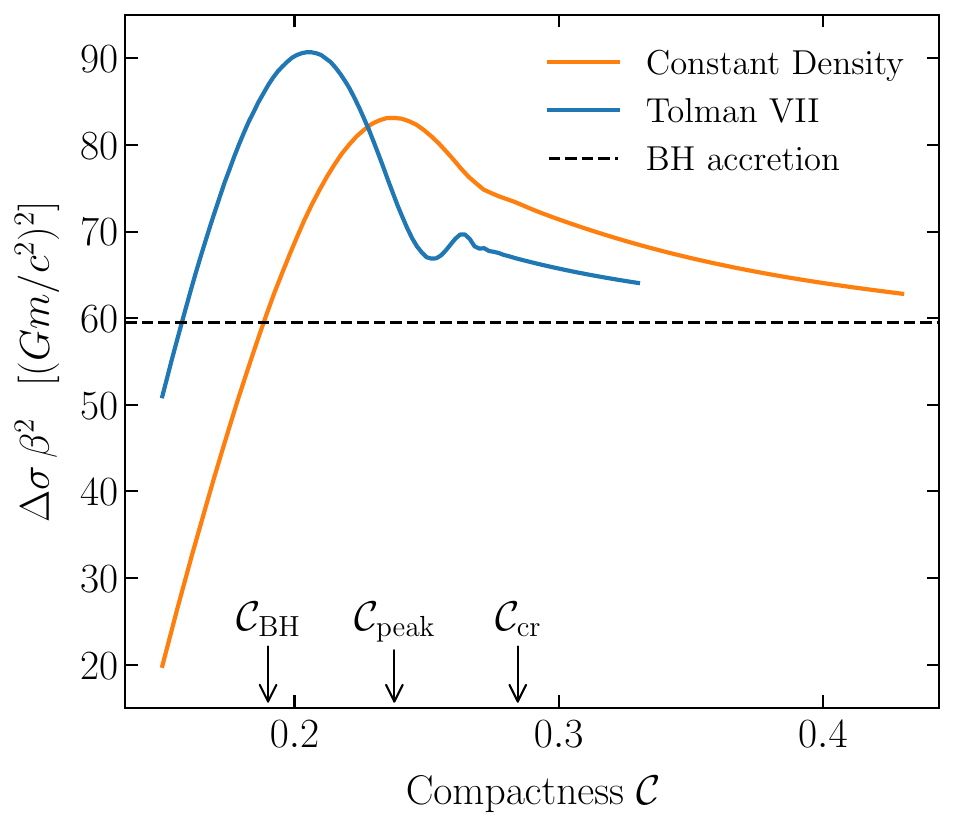}
    \caption{\textbf{Difference in scattering cross-section for an extended toy object and a point mass.} The dashed black line represents the accretion cross-section for a BH. \label{fig:dcsection_toy}}
\end{figure}

To demonstrate the robustness of the observed trends, we reproduce our numerical results using idealized stellar models: a constant-density sphere and a Tolman–VII configuration \cite{PhysRev.55.364}. As is evident for the geometric unit representation of these models, they depend primarily on the compactness, with the total mass only scaling the cross-section quadratically. For realistic equations of state, and in the absence of twin-star branches \cite{2000A&A...353L...9G}, fixing the mass effectively fixes the radius and therefore $\mathcal{C}$. For concreteness, we evaluate the cross-section at the average orbital velocity of the radial snapshot in \cref{fig:force_compactness}, which provides a representative dynamical scale for the encounters. As shown in \cref{fig:dcsection_toy}, the toy-models reproduce all the characteristic features of \cref{fig:force_compactness}. Additionally, we observe that for higher values of compactness, the scattering cross section appears to asymptotically mimic BH accretion, though we only report the behavior up to each model's stability limit.

For reference, we define $\mathcal{C}_\mathrm{BH}$ and $\mathcal{C}_\mathrm{peak}$ through the constant density model for which we report the following numerical fits
\begin{align}
    \mathcal{C}_\mathrm{peak} &\approx 0.097 +0.48 \beta -0.47\beta^2 +0.2\beta^3\,,\\
    \mathcal{C}_\mathrm{BH} &\approx 0.04 +0.5 \beta -0.47\beta^2 +0.2\beta^3\,,
\end{align}
that were used to mark their respective points in \cref{fig:force_compactness}. For completeness, the critical compactness associated with plunging trajectories is given by $\mathcal{C}_\mathrm{cr} = Gm/c^2/r_0^\mathrm{min}$. In general, because of NS's crust reduces the stellar compactness, these predictions slightly overestimate the true values for realistic EOS.\\

\paragraph*{\bf Analytical understanding.}
Motivated by the robustness of these trends across numerical experiments and toy models, we next develop a minimal analytical description to capture their physical origin. First, the cross-section decline at low compactness is easy to interpret: as $\mathcal{C}$ approaches $0$, the object becomes infinitely dilute and its dynamical friction vanishes, and $\lim_{\mathcal{C}\to 0}\delta\sigma = -(\sigma_\mathrm{BH} +\sigma_\mathrm{acc})$. For the rest, it is useful to recast \cref{eq:dsigma} in terms of the deflection angle,
\begin{equation} \label{eq:delta_sigma_approx}
    \delta\sigma \approx \sigma_\mathrm{acc} - \pi\int_0^{b_\mathrm{cr}} \!\!\! \cos\theta_\mathrm{NS} \,\mathrm{d}b^2\,,
\end{equation}
where we have neglected a third term integrated in $b \in (b_\mathrm{cr}, b_\mathrm{th})$ which is either small or zero. Evidently, the signature of the remaining integral in \cref{eq:delta_sigma_approx} directly determines the force hierarchy between a NS and a BH. In practice, if most of the radially infalling particles that would be accreted by a BH are instead strongly deflected by the NS on obtuse angles, the NS cross-section becomes larger. This is possible with increasing stellar compactness where deflection grows stronger, as particles probe increasingly extreme curvature just outside the NS. It also follows that, beyond a certain compactness, $\mathcal{C}_\mathrm{peak}$, particles begin to wind around the star, reducing the net deflection and thereby diminishing the dynamical friction, leading to the observed peak. Finally, for extremely compact objects, particles with $b \lesssim b_\mathrm{cr}$ become effectively trapped, winding around the star for many cycles. Then, the corresponding integral averages out through rapid oscillations, leading to $\delta\sigma \approx \sigma_\mathrm{acc}$ as the compactness increases.

\begin{figure*}[ht!]
\centering
\hspace*{-0.08\columnwidth}
    \includegraphics[width=0.48\textwidth]{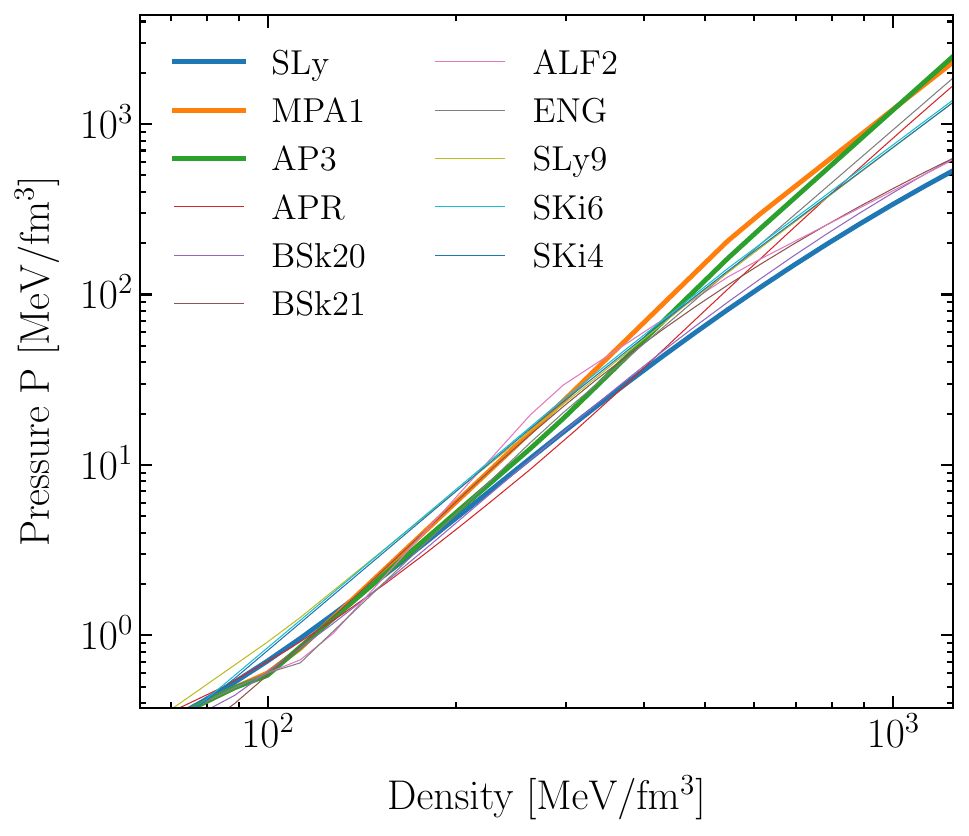}
    \includegraphics[width=0.49\textwidth]{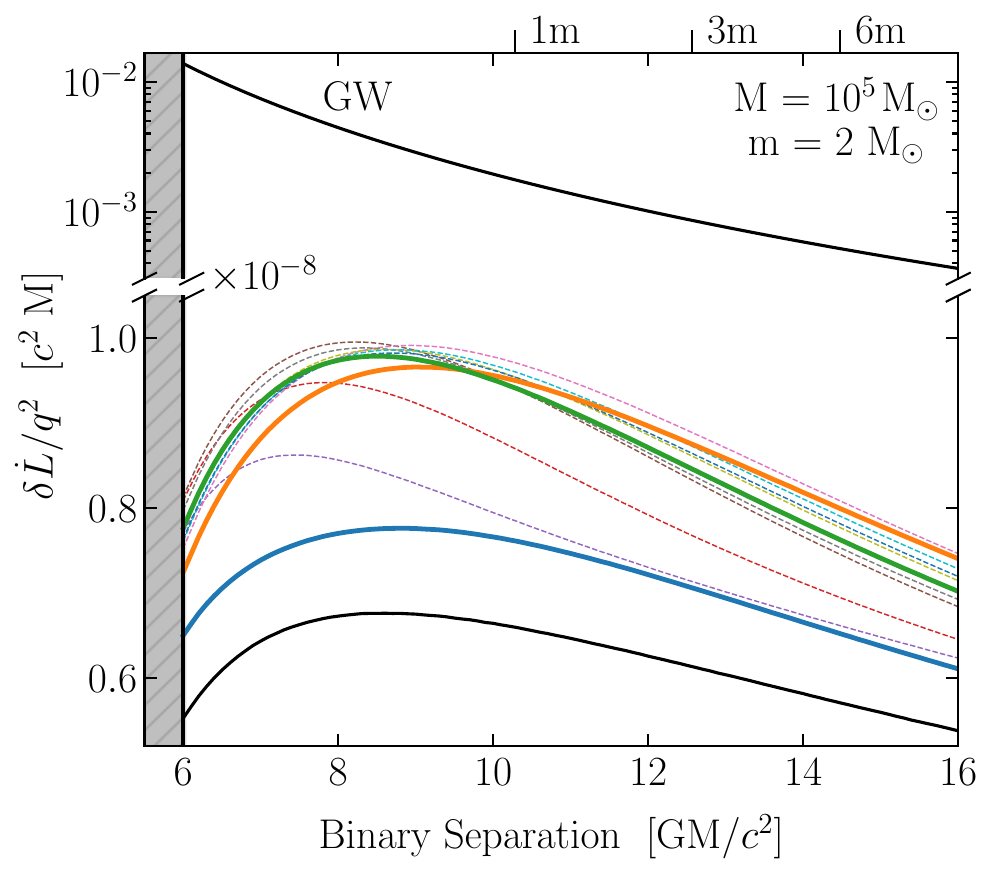}
    \caption{\textbf{Equations of state (left) and angular-momentum loss of neutron stars 
    as a function of binary separation (right).} For comparison, we show the corresponding accretion-driven loss for a BH of the same mass (black), as well as the loss from gravitational-wave emission, which is plotted on a disconnected y-axis on top. Along the top axis we indicate the initial separations at which a binary would plunge after the corresponding number of months. All results assume the benchmark dark matter spike for a system with a $\mathrm{M}=10^5~\mathrm{M}_\odot$ primary, representative of the density inferred for Draco dSph \cite{Wanders_2015}. Aside from the three fiducial equations of state, we also show a collection of 8 other representative models: (AP4 \cite{Akmal_1998}, BSk20-21 \cite{Goriely_2010}, ALF2 \cite{Alford_2005}, ENG \cite{1996ApJ...469..794E}, SLy9 \cite{Suleiman_2022}, SKi6 \cite{PhysRevC.53.740}, SKi4 \cite{REINHARD1995467}) using unified \cite{Potekhin_2013,Goriely_2010,PhysRevC.85.065803} and polytrope \cite{Read_2009,Suleiman_2022} modelling with a crust as in \cite{Douchin_2001}.\label{fig:equations_of_state}}
\end{figure*}

\section{The Bayesian framework}
\label{app:bayesian}

To assess the prospects of measuring the equation of state of a NS companion in an EMRI we follow the Bayesian approach employed in Refs.~\cite{Coogan_2022,Cole:2022yzw,Cole_2023}. For simplicity we study face-on binaries and set the viewing angle to $0$. We assume the strain time-series measured by the detector $d(t)$ to be the sum of the signal $s(t)$ and the detector's noise $n(t)$ which we take to be Gaussian to write the likelihood function
\begin{equation}
    p(d|h_{\bm{\theta}}) \propto \exp\Big[ \langle h_{\bm{\theta}} |d\rangle -  \langle h_{\bm{\theta}} |h_{\bm{\theta}}\rangle\Big]\,.
\end{equation}
Then, the terms $\left< a | b \right>$ are the overlaps of two waveforms $a$ and $b$ in the frequency domain defined as
\begin{equation}
    \left<a| b\right> \equiv 4 \,\mathrm{Re} \int_{f_\mathrm{min}}^{\infty} \frac{a^*\left(f\right)b\!\left(f\right)}{S_n\left(f\right)} \, \mathrm{d}f \,, \label{eq:overlap_integral}
\end{equation}
with $S_n\left(f\right)$ taken to be the one-sided LISA power-spectral density taken to be the Michelson-like PSD from~\cite{Robson_2019}, and\footnote{We note that changes in the slope of the cusp are mostly degenerate with the density normalization and preliminary analysis shows that they don't affect the shape of the losses within the regions we are interested in.} 
\begin{equation*}
    \bm{\theta} = \{ M, m, \rho, d_L, \phi_c, t_c\}
\end{equation*}
are the parameters describing the model waveform. The binary masses $M$, $m$ and a factor describing the spike's density normalization $\rho$ are intrinsic parameters for which the likelihood is maximized over numerically. Conversely, the binary's luminosity distance $d_L$, and the waveform's phase and time offsets $\phi_c$, $t_c$ can all be maximized analytically, to yield the new likelihood,
\begin{equation}
    p_\mathrm{max}\left(d | h_{\bm{\theta}}\right) \equiv \exp\left[ \frac{\left<h_{\bm{\theta}} | d \right>^2_\mathrm{max}}{2 \left< h_{\bm{\theta}} | h_{\bm{\theta}}\right>} \right] \,,
\end{equation}
where with the subscript ``$\mathrm{max}$'' we indicate maximization over phase and time offsets of the signal as described in \cite{Coogan_2022}. Finally, the EMRI waveforms $h_{\bm\theta}, s$ are obtained by incorporating the angular momentum rate from \cref{eq:angtum}[Eq. 1 of the paper] for each EOS into the \texttt{FastEMRIWaveforms\_v2.0.0} (FEW) framework \cite{Katz:2021yft,Chua_2021,Speri_2024,chapmanbird2025fastframedraggingefficientwaveforms} framework, and initializing binaries 1 year before merger. Because we will be working with large SNR $\sim \sqrt{\langle d|d\rangle}$, we take $d(t) \approx s(t)$. The effective sample size in our experiments was approximately 1500, which is sufficient for robust statistical inference \cite{ALBERT2025109756}.

The posterior distribution $p({\bm\theta}|d) \propto p_\mathrm{max}(d | h_{\bm\theta}) \, p({\bm\theta})$ is explored using \textit{static} nested sampling \cite{Higson_2018,2004AIPC..735..395S,10.1214/06-BA127} implemented with \texttt{dynesty\_v2.1.5} \cite{2020MNRAS.493.3132S,sergey_koposov_2025_17268284}. This algorithm is primarily optimized for robust evidence estimation rather than maximal posterior resolution. As a result, the posterior contours may appear less smooth, although this does not affect the accuracy of the evidence calculation. Mapping out these posteriors allows us to assess how well the model's parameters can be measured and whether bias is induced by performing inference with the wrong model.

The prior distributions $p(\theta)$ are taken to be uniform $\mathcal{U}$ accordingly: $\rho = \rho_0 \, \mathcal{U}\left(9, 11\right)$, $M = M_0 +\mathcal{U}\left(-15, 15 \right) \mathrm{M}_\odot$, $m = m_0 +10^{-5}  \, \mathcal{U}\left( -15, 15\right) \mathrm{M}_\odot$, where $\rho_0, M_0, m_0$ are the values of the injected parameters. These are relatively narrow to mitigate the computational cost of EMRI inference \cite{cole2025sequentialsimulationbasedinferenceextreme,PhysRevD.108.084014} and isolate the effect of the internal structure. 

Finally, to assess the ability to distinguish between different companion models we compute the evidence for each model/hypothesis
\begin{equation}
    p(d) = \int p_\mathrm{max}(d | h_{\bm\theta}) \, p({\bm\theta}) \, \mathrm{d}{\bm\theta}\,,
\end{equation}
and compare them in competing pairs $H_k$ using the Bayes factor \cite{Jeffreys1939-JEFTOP-5}, defined as the ratio
\begin{equation}
    \mathcal{B} \equiv \frac{p(d|H_1)}{p(d|H_2)}\,.
\end{equation}
A large Bayes factor denotes a confident Bayesian preference for model $H_1$ over $H_2$.

Alternatively, a useful approximation for estimating the Bayes factor, is the Laplace approximation $\ln \mathcal{B} = \mathrm{SNR}^2 \left( 1 -\mathcal{F}^2 \right)\!/2$ \cite{PhysRevD.103.062002,Cornish_2011} where $\mathcal{F}$ is the \textit{fitting factor} calculated by maximizing \cref{eq:overlap_integral} over $\bm{\theta}$ and normalizing it such that $\mathcal{F} = 1$ for the same waveform.

\section{A parametrised measurement of the stellar radius}
We now consider whether it is possible to measure properties of the NS structure, with minimal model assumptions. This approach allows constraints on the EOS without fixing it a priori. To this end, we approximate the space-time within the NS as that sourced by a generalized super-ellipsoid (``Gensel'') density profile of the form,
\begin{equation} \label{eq:gensel}
    \rho(r) = \rho_c \left(1 -\frac{r^2}{R^2} \right)^a\,.
\end{equation}

This family of profiles generalizes the well-known, fully analytic Tolman VII solution \cite{PhysRev.55.364}, which has been shown to approximate numerical solutions for a wide range of EOS models \cite{Lattimer_2001,Raghoonundun_2015}. Within this framework, the parameter $a$ regulates the steepness with which the density tapers off towards the stellar radius, thus effectively controlling the relative size between NS core and crust: larger values of $a$ produce thicker crusts. For $a =0$, the model reduces to a constant density profile sourcing the interior Schwarzschild metric, and for $a = 1$, it reduces to the Tolman VII solution \cite{Jiang_2019}.

We find the mass enclosed within radius $r$ to be given analytically by,
\begin{equation} \label{eq:modified_mr}
    m_r(r) = \frac{4\pi \rho_c r^3}{3} \, {}_2F_1\left(-a,\, 3/2,\, 5/2,\, r^2/R^2\right)\,,
\end{equation}\\
where it is useful to eliminate the central density $\rho_c$ from the expressions by substituting it with the total mass,
\begin{equation}
    M = \frac{4\pi \rho_c R^3}{3} \frac{\Gamma(5/2)\Gamma(a+1)}{\Gamma(a+5/2)}\,.
\end{equation}
It is straightforward to obtain a closed form expression for $g_\mathrm{rr}$ by substituting \cref{eq:modified_mr} into \cref{eq:ns_metric}[Eq. A2], but we must integrate \cref{eq:dlngtt}[Eq. A3] numerically for $g_\mathrm{tt}$ after computing $P(r)$ imposing $P(R) = 0$.\footnote{We omit closed-form expressions for $P(r)$ and $g_\mathrm{tt}(r)$ as they are algebraically unwieldy.} We then proceed to solve the geodesic motion of particles within our modified model to compute dynamical friction and produce relativistic waveforms for inspirals within DM spikes.

\begin{figure}[ht!]
\centering
\hspace*{-0.08\columnwidth}
    \includegraphics[width=\columnwidth]{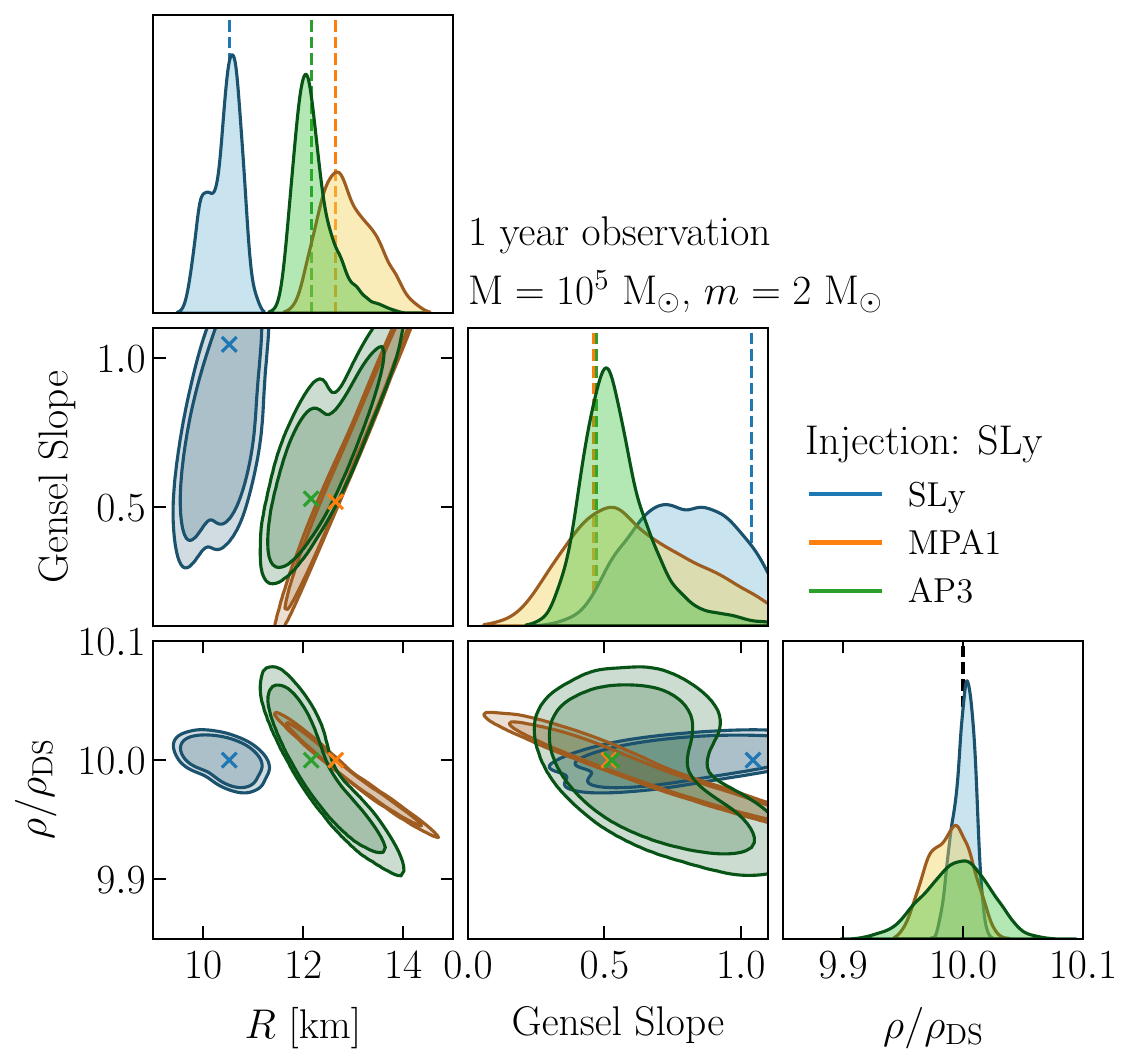}
    \caption{\textbf{Marginal posteriors for the parameterized neutron star profile and the spike's density.} Dashed lines and crosses indicate true values with $a$ obtained by fitting \cref{eq:gensel} to each density profile}. \label{fig:radius_corner}
    \vspace*{2em}
\end{figure}

In \cref{fig:radius_corner}, we show the three-dimensional marginal posteriors of the NS radius, gensel slope parameter, and DM density.
The posteriors are obtained by injecting the three fiducial NS systems and recovering them with our modified model. For simplicity, and given the small inferred bias in primary and secondary mass (cf. \cref{fig:corern_ns}[Fig. 2 of the paper]), we fix these two parameters in this calculation to their true value ($\mathrm{M} = 10^5~\mathrm{M}_\odot$ and $m = 2~\mathrm{M}_\odot$). We find that the radius can be recovered with an uncertainty of $\pm~700~\mathrm{m}$ in all cases, and that the tapering strength of the profile can also be inferred, allowing discrimination between small and large values of $a$. A more thorough investigation of this approach is deferred to future work.

\bibliography{bibliography}

\end{document}